\documentclass[pra,twocolumn,superscriptaddress,10pt,noshowpacs]{revtex4}
\usepackage[english]{babel}
\usepackage[T1]{fontenc}
\usepackage[utf8]{inputenc}
\usepackage{graphicx,epstopdf}
\usepackage{amsmath}

\usepackage{amsfonts}
\usepackage{bbm}
\usepackage{amssymb}
\usepackage{color}
\usepackage{latexsym}
\usepackage{caption}
\usepackage{subcaption}
\usepackage{times,txfonts}

\begin{document}

\title{Black string bounce to traversable wormhole}

\author{A. Lima}
\affiliation{Universidade Federal do Cear\'{a}, Fortaleza, Cear\'{a}, Brazil}
\email{arthur.lima@fisica.ufc.br}

\author{G. Alencar}
\affiliation{Universidade Federal do Cear\'{a}, Fortaleza, Cear\'{a}, Brazil}
\email{geova@fisica.ufc.br}

\author{J. Furtado}
\affiliation{Universidade Federal do Cariri(UFCA), Av. Tenente Raimundo Rocha, \\ Cidade Universit\'{a}ria, Juazeiro do Norte, Cear\'{a}, CEP 63048-080, Brasil}
\email{job.furtado@ufca.edu.br}

\date{\today}

\begin{abstract}
In this work, a regular black string solution will be presented from the method used by Simpson-Visser to regularize the Schwarzschild solution. As in the Simpson-Visser work, in this new black string solution, it is possible to represent both a regular black hole and a wormhole just by changing the value of a parameter ``$a$'' used in its metric. Tensors and curvature invariants were analyzed to verify the regularity of the solution, as well as the energy conditions of the system. It was found that the null energy condition will always be violated for the entire space. The analysis of the thermodynamic properties of the regular black string was also carried out, in which the modifications generated about the original solution of the black string, were evaluated, specifically, the Hawking temperature, entropy, its thermal capacity, and the Helmholtz free energy. Finally, we investigate the possible stable or unstable circular orbits for photons and massive particles. The results were compared with those of the non-regular black string, seeking to make a parallel with the Simpson-Visser work.  
\end{abstract}

\maketitle

\section{Introduction}\label{Intro}

\hspace{0.4cm}Black holes are solutions of Einstein's equations in which there is the presence of a surface called the event horizon in which any matter or even light that passes through it can only describe a one-way path, not even light can travel off the horizon without violating causality \cite{visser1995lorentzian}. This type of solution is very important in Relativity, as it can define the type of geometry created when the body collapses, such as a star or star cluster \cite{Lemos:1994xp}. These solutions have gained a lot of relevance in recent years due to technological evolution, which can be evidenced by the detection of gravitational waves by LIGO and VIRGO, and by the first images of supermassive black holes \cite{LIGOScientific:2016aoc, EventHorizonTelescope:2022wkp, EventHorizonTelescope:2019dse, LIGOScientific:2017vwq}.

In General Relativity, the main solutions for black holes form a family of 4 basic parameters called the generalized Kerr-Newman family. The 4 parameters are the mass $M$, the angular momentum $J$, the electric charge $Q$, and the cosmological constant $\Lambda$. All these solutions have axial symmetry (axisymmetric), and can be asymptotically flat (when the cosmological constant is zero), \textit{de Sitter} (if $\Lambda>0$), or \textit{anti-de Sitter} (if $\Lambda<0$), so the asymptotic behavior depends directly on the cosmological constant \cite{Lemos:1994xp}. A possible, less common solution, is to consider cylindrical symmetry. This type of solution is particularly important for cosmology, because in the study of the evolution of the Universe, in the phase transitions that may have occurred after the \textit{big bang}, the so-called topologically stable defects are studied, such as cosmic strings, which generate very interesting results such as the density fluctuation that explains the creation of galaxies \cite{Vilenkin:1984ib}. These cosmic strings can be worked with cylindrical symmetry for spacetime. We also call the black string cylindrical black hole solution. 

Basic black hole solutions to spacetime in General Relativity have a singularity at their origin since matter collapses into it. This is usually ignored, as there is not much physical interest in analyzing the origin of a black hole. However, when we are working with black hole evaporation, these singularities cannot be ignored in the final stages of evaporation. \cite{Carballo-Rubio:2018pmi, Jayawiguna:2022tly}. However, it is possible to develop regular solutions for black holes. Spherically symmetric, static, asymptotically flat, and regular metrics can be found, including being physically reasonable, without violating the energy conditions \cite{Hayward:2006}.

In this context, Simpson and Visser \cite{Simpson:2018tsi} proposed a regular solution through a modification of the Schwarzschild solution, presenting the following metric as a candidate to generate this solution:

\begin{equation}\label{1}
    ds^2=-\left(1-\frac{2m}{\sqrt{r^2+a^2}}\right)dt^2+\frac{dr^2}{\left(1-\frac{2m}{\sqrt{r^2+a^2}}\right)}+(r^2+a^2)d\Omega^2,
\end{equation}
where ``$a$'' represents an adjustable parameter, and ``$m$'' is a constant that is related to the mass of matter that will generate this curved spacetime. Basically, the change $r^2\rightarrow r^2+a^2$ was carried out in the Schwarzschild solution. In summary, this metric has the following properties: if $a>2m$ we have a two-way Morris-Thorne wormhole \cite{Simpson:2018tsi, Morris:1988cz}; when $a=2m$ we have a one-way wormhole with a horizon in the ``throat'', similar to a Schwarzschild wormhole \cite{Simpson:2018tsi, Morris:1988cz}; and if $a<2m$ we have a regular one-way black hole with event horizons located at $r_H=\pm \sqrt{(2m)^2-a^2}$ \cite{Simpson:2018tsi}. After an analysis of the tensors and curvature invariants, it is verified that there are no singularities at the origin since $a\neq 0$. Another important fact is the violation of energy conditions due to the presence of matter to generate this solution, at any point in space-time and regardless of whether we have a black hole or a wormhole.


In addition, Simpson and Visser found expressions for the Hawking temperature of the regular solution and the circular orbits for photons (sphere of photons) and massive particles (ISCO), finding results similar to those of the Schwarzschild solution, but with a correction factor of the type $\sqrt{1-a^2/k^2}$, where $k$ is a constant multiple of $m$. In the case of temperature, $k=2m$ \cite{Simpson:2018tsi}.

After the publication of this work, several others appeared along the same lines of treating an already known non-regular solution and performing the Simpson-Visser regularization. Simpson himself published a work with the same regularization for Reisnerr-Nordstrom and Kerr-Newman black holes \cite{Simpson:2021vxo}, thin disc accretion study for Simpson-Visser solution can be found at \cite{Bambhaniya:2021ugr}, can be used this solution to verify the impossibility of the existence of traversable wormholes in semiclassical gravitation \cite{Terno:2022qot}, in addition to works with modified gravity \cite{Junior:2022fgu}, gravitational lens \cite{Islam:2021ful, Nascimento:2020ime, Tsukamoto:2020bjm}, with phantom fields \cite{Chataignier:2022yic, Bronnikov:2022bud}, among many others.

However, up to now, there is a lack in the literature regarding the regularization of the solution of a black string, which is the purpose of this work. This work is organized as follows: In the next section we perform the Simpson-Visser regularization in the metric obtained by Lemos in order to analyze if it occurs bounce from black string to a wormhole. We compute also the tensors and curvature invariants in order to investigate their regularity. Moreover we study the energy conditions of this solution. In section III we perform an analysis of its thermodynamic properties, and we study the possible circular orbits for photons and massive particles. In section IV we draw our conclusions.   

\section{Simpson-Visser regularization for a black string}\label{Sec-2}
Let us briefly review the exact solution of a cylindrically symmetrical black hole with a negative cosmological constant in the context of general relativity, also called a black string. Thus we are considering the coordinates $x^{\mu}=(t,\, r,\, \phi,\, z)$. These coordinates vary according to the following ranges: $t\in (-\infty,\, \infty);\, r\in (-\infty,\, \infty);\, \phi \in [0,\, 2\pi];\, z\in (-\infty,\, \infty)$. For this type of solution, it is necessary to have a negative cosmological constant \cite{Lemos:1994xp}. 

The ansatz for the metric is given by

\begin{equation}\label{3}
    ds^2=-f(r)dt^2+\frac{dr^2}{f(r)}+r^2d\phi^2+\alpha^2r^2dz^2,
\end{equation}
where $\alpha^2\equiv -\frac{1}{3}\Lambda>0$ \cite{Lemos:1994xp}. For our purposes, it is only necessary to use one component of the Einstein tensor, the component $G^0{}_0$. We will, therefore, obtain the following expression for $G^{0}{}_{0}$

\begin{equation}\label{4}
    G^{0}{}_{0} = \frac{[f(r)+rf'(r)]}{r^2}=\frac{1}{r^2}\frac{d}{dr}[rf(r)],
\end{equation}
where $f'(r)=df(r)/dr$. We are interested in the region where $r\neq 0$, which allows us to assume that, in this region, $T^{0}{}_{0}=0$. So, using Einstein's equation for $G^0{}_0$ and $T^{0}{}_{0}$, we have

\begin{equation}\label{5}
    \frac{1}{r^2}\frac{d}{dr}[rf(r)] + \Lambda = 0,
\end{equation}
whose solution for $f(r)$ is given by
\begin{equation}\label{6}
     f(r) = -\frac{\Lambda}{3}r^2+\frac{C}{r},
\end{equation}
where $C$ is a constant of integration. Following the usual solution defined in \cite{Lemos:1994xp}, we have that

\begin{equation}\label{7}
     f(r) = \alpha^2r^2-\frac{b}{\alpha r}.
\end{equation}
The event horizon can be determined simply by doing $g_{00}=0$

\begin{equation}\label{8}
    r \equiv r_{HL} = \frac{b^{1/3}}{\alpha}.
\end{equation}
This solution is not regular, as the Kretschmann scalar $R^{\mu\nu\lambda\rho}R_{\mu\nu\lambda\rho} = 24\alpha^4\left(1+\frac{b ^2}{2\alpha^6r^6}\right)$ has singularity at $r=0$. Another interesting result is that the Ricci scalar will be constant and proportional to the cosmological constant, specifically, $R=4\Lambda=-12\alpha^2$, indicating that, in the asymptotic limit, the curvature will not be zero. Therefore, the solution is not asymptotically flat.

Now, following the Simpson-Visser regularization scheme of the Schwarzschild metric, we seek to perform the same kind of regularization for the black string solution. 
In order to perform the regularization, we must consider the following transformation $r\rightarrow\sqrt{r^2+a^2}$ in the metric defined in (\ref{3}) with $f(r)$ given by (\ref{7}), which yields

\begin{eqnarray}\label{9}
   \nonumber ds^2&=&-\left(\alpha^2(r^2+a^2)-\frac{b}{\alpha \sqrt{r^2+a^2}}\right)dt^2+\\
   && \frac{dr^2}{\left(\alpha^2(r^2+a^2)-\frac{b}{\alpha \sqrt{r^2+a^2}}\right)}+(r^2+a^2)[d\phi^2+\alpha^2dz^2]. 
\end{eqnarray}
   
For simplicity's sake, here we'll make another change to the radial coordinate: $\overline{r}^2=r^2+a^2$. However it should be noticed that here we are not changing the coordinates to return to the same Lemos solution, we are making a coordinate change, which means that we will continue with the same solution defined in (\ref{9}). Now it will take on a simple and more intuitive form as

\begin{eqnarray}\label{10}
   \nonumber ds^2&=&-\left(\alpha^2\overline{r}^2-\frac{b}{\alpha \overline{r}}\right)dt^2+\frac{d\overline{r}^2}{\left(1-\frac{a^2}{\overline{r}^2}\right)\left(\alpha^2\overline{r}^2-\frac{b}{\alpha \overline{r}}\right)}+\\
   && \overline{r}^2d\phi^2+\alpha^2\overline{r}^2dz^2.
\end{eqnarray}
The above metric can be rewritten as a function of the event horizon position $\overline{r}_H=b^{1/3}/\alpha$, so that,

\begin{eqnarray}\label{11}
  \nonumber ds^2&=&-\alpha^2\overline{r}^2\left(1-\frac{\overline{r}_H^3}{\overline{r}^3} \right)dt^2+\frac{d\overline{r}^2}{\alpha^2\overline{r}^2\left(1-\frac{a^2}{\overline{r}^2}\right)\left(1-\frac{\overline{r}_H^3}{\overline{r}^3}\right)}+\\ && \overline{r}^2d\phi^2+\alpha^2\overline{r}^2dz^2.
\end{eqnarray}

Clearly, by setting $a=0$ we recover the Lemos solution, which is not regular, so, for our interest, we must assume $a\neq 0$. Let us perform an analysis of null radial curves, that can be achieved by considering $ds^2=d\phi=dz=0$, leading to

\begin{equation}\label{12}
    \frac{d\overline{r}}{dt}=\pm \alpha^2\overline{r}^2\left(1-\frac{\overline{r}_H^3}{\overline{r}^3}\right) \sqrt{1-\frac{a^2}{\overline{r}^2}}.
\end{equation}

As $\overline{r}=\pm \sqrt{r^2+a^2}$, for $r=0$, we will have $\overline{r}=\pm a$, so $a$ is the minimum value that the radial coordinate can have, similar to the ``throat'' of a wormhole. If $a>\overline{r}_H$, like $\overline{r}\geq a$, the event horizon will never be reached, that is, the temporal term of the metric will never vanish, which implies that $d \overline{r}/dt$ will always be non-zero, except for $\overline{r}=a$, but this value is outside the event horizon. Therefore, this solution is characteristic of a traversable wormhole, but it is not of the Morris-Thorne type, since its symmetry is not spherical, but cylindrical. If $a=\overline{r}_H$, it means that in $\overline{r}=a$ we will have an event horizon, exactly at the ``throat'' of the wormhole, similar to the Simpson-Visser example. So we're going to have a wormhole with the extreme ``throat'' of one-way motion. Finally, for $a<\overline{r}_H$, the event horizon can be reached at a value different from the minimum value $a$, as this has a value smaller than the horizon and the coordinate $\overline{ r}$ can assume any value greater than or equal to $a$. Consequently, as the coordinate moves away from the minimum value that is $a$ itself, it will approach and reach the horizon, thus indicating a black hole solution. 

Now we need to compute the tensors and the curvature invariants to analyze the regularity or not of these solutions and finally to study the Einstein equations and the energy conditions.

\subsection{\textit{Tensors and curvature invariants}}
\label{TIC}
Our job now is to determine the Riemann and Ricci tensors, as well as the Ricci, Ricci contraction, and Kretschmann scalars from the metric defined in the equation (\ref{11}). The non-zero components of the Riemann tensor are

\begin{eqnarray}
    R^{01}{}{}_{01} &=& \frac{\alpha^2(2\overline{r}_H^3\overline{r}^2-3a^2\overline{r}_H^3-2\overline{r}^5)}{2 \overline{r}^5}; \label{13}\\
    R^{02}{}{}_{02}&=&R^{03}{}{}_{03}=\frac{\alpha^2(a^2-\overline{r}^2)(2\overline{r}^3+\overline{r}_H^3)}{2 \overline{r}^5}; \label{14}\\
    R^{12}{}{}_{12}&=&R^{13}{}{}_{13}=\frac{\alpha^2(3a^2\overline{r}_H^3-\overline{r}_H^3\overline{r}^2-2\overline{r}^5)}{2 \overline{r}^5}; \label{15}\\
    R^{23}{}{}_{23}&=&\frac{\alpha^2(a^2-\overline{r}^2)(\overline{r}^3-\overline{r}_H^3)}{ \overline{r}^5}. \label{16}
\end{eqnarray}

The non-zero components of the Ricci tensor can now be determined and they are given by

\begin{eqnarray}
    R^{0}{}_{0} &=& \frac{\alpha^2[a^2(4\overline{r}^3-\overline{r}_H^3)-6\overline{r}^5]}{2 \overline{r}^5}; \label{17}\\
    R^{1}{}_{1} &=& \frac{\alpha^2(3a^2\overline{r}_H^3-6\overline{r}^5)}{2 \overline{r}^5};\label{18}\\
    R^{2}{}_{2} &=& R^{3}{}_{3} = \frac{a^2\alpha^2}{\overline{r}^2}\left(\frac{\overline{r}_H^3}{ \overline{r}^3}+2\right)-3\alpha^2. \label{19}
\end{eqnarray}
Adding all these components together we can finally compute the Ricci scalar

\begin{equation}\label{20}
    R = -12 \alpha^2 + \frac{3a^2\alpha^2(\overline{r}_H^3+2\overline{r}^3)}{ \overline{r}^5}.
\end{equation}
With the components of the Ricci tensor and with the Ricci scalar, we can determine the non-zero components of the Einstein tensor, which are

\begin{eqnarray}
    G^{0}{}_{0} &=& \frac{\alpha^2[3\overline{r}^5-a^2(\overline{r}^3+2\overline{r}_H^3)]}{\overline{r}^5}; \label{21}\\
    G^{1}{}_{1} &=& \frac{3\alpha^2(\overline{r}^2-a^2)}{\overline{r}^2}; \label{22}\\
    G^{2}{}_{2} &=& G^{3}{}_{3} = 3\alpha^2-\frac{a^2\alpha^2}{\overline{r}^2}\left(\frac{\overline{r}_H^3}{2 \overline{r}^3}+1\right). \label{23}
\end{eqnarray}
Finally, the Ricci contraction and the Kretschmann scalar are given by

\begin{eqnarray}
    \nonumber &&R^{\mu\nu}R_{\mu\nu} = 36 \alpha^4 - \frac{18a^2\alpha^4(\overline{r}_H^3+2\overline{r}^3)}{\overline{r}^5} + \\ && \frac{3a^4\alpha^4(3\overline{r}_H^6+4\overline{r}_H^3\overline{r}^3+8\overline{r}^6)}{2\overline{r}^{10}}; \label{24}\\
     \nonumber &&R^{\mu\nu\lambda\rho}R_{\mu\nu\lambda\rho} = \frac{12\alpha^4(\overline{r}_H^6+2\overline{r}^6)}{\overline{r}^6}+\\ && \frac{3\alpha^4[a^4(11\overline{r}_H^6+4\overline{r}^6)-4a^2(3\overline{r}_H^6\overline{r}^2+\overline{r}_H^3\overline{r}^5+2\overline{r}^8)]}{\overline{r}^{10}}. \label{25}
\end{eqnarray}
Analyzing the components of the Riemann tensor and the scalars of curvature in the minimum value of $\overline{r}$ which is $a$, we will obtain the following results:

\begin{eqnarray}
    &&R^{01}{}{}_{01} = -\frac{\alpha^2(\overline{r}_H^3+2a^3)}{2 a^3}; R^{02}{}{}_{02}=R^{03}{}{}_{03}=R^{23}{}{}_{23}=0; \label{26}\\
    &&R^{12}{}{}_{12}=R^{13}{}{}_{13}=\frac{\alpha^2(\overline{r}_H^3-a^3)}{a^3}; \label{27}\\
    &&R = -12 \alpha^2 + \frac{3\alpha^2(\overline{r}_H^3+2a^3)}{a^3}; \label{28}\\
    &&R^{\mu\nu}R_{\mu\nu} = 12 \alpha^4 - \frac{12\alpha^4 \overline{r}_H^3}{a^3} + \frac{9\alpha^4\overline{r}_H^6}{2a^{6}}; \label{29}\\
    &&R^{\mu\nu\lambda\rho}R_{\mu\nu\lambda\rho} = \frac{3\alpha^4(3\overline{r}_H^6+4a^6-4\overline{r}_H^3a^3)}{a^{6}}. \label{30}
\end{eqnarray}
Therefore, as $a\neq 0$, we see that there are no singularities in the curvature of this spacetime, that is, we have a regular solution for any non-zero parameter $a$. However, this solution is a little different from the one found by Simpson-Visser, because its metric is not asymptotically flat, which can be seen when we do the same analysis, but with $|r|\rightarrow \infty$, since all the components of the Riemann tensor will tend to $-\alpha^2$, the Ricci scalar will tend to $-12\alpha^2$, the Ricci contraction to $36\alpha^4$ and the Kretschmann scalar to $24 \alpha^4$, that is, in the asymptotic limit the curvature is not zero and will depend on the cosmological constant. 

\subsection{Black string moment-energy tensor and energy conditions}
We will now discuss the energy conditions for the Simpson-Visser modified black string. We can define the pressures and energy density through the energy-momentum tensor as follows

\begin{equation}\label{31}
    T^0{}_0=-\rho;\, T^{1}{}_{1}=p_{\|};\, T^{2}{}_{2}=T^{3}{}_{3}=p_{\bot}.
\end{equation}
At first we should define $T^{2}{}_{2}=p_{\phi}$ and $T^{3}{}_{3}=p_{z}$, but since $G^{ 2}{}_{2}=G^{3}{}_{3}$, we have $T^{2}{}_{2}=T^{3}{}_{3}$. 

Using Einstein's equations with cosmological constant, we can obtain the expressions for the components of the energy-momentum tensor as

\begin{eqnarray}
    \rho &=& \frac{\alpha^2a^2(\overline{r}^3+2\overline{r}_H^3)}{8\pi G \overline{r}^5}; \label{32}\\
    p_{\|} &=& -\frac{3\alpha^2a^2}{8\pi G \overline{r}^2}; \label{33}\\
    p_{\bot} &=& - \frac{a^2\alpha^2(2\overline{r}^3+\overline{r}_H^3)}{16\pi G \overline{r}^5}. \label{34}
\end{eqnarray}
To assess whether this solution satisfies the energy conditions, in particular, the null energy condition ($\rho+p_i\geq 0$). So let us calculate $\rho+p_{\|}$

\begin{equation}\label{35}
    \rho + p_{\|} = -\frac{2a^2\alpha^2(\overline{r}^3-\overline{r}_H^3)}{8\pi G \overline{r}^5}, 
\end{equation}
which is a valid result outside the event horizon, i.e., for $\overline{r}>\overline{r}_H$. We see, therefore, that the right-hand-side of the equation (\ref{35}) will always be negative. This implies $\rho+p_{\|}<0$, that is, the null energy condition is violated, similarly to the Simpson-Visser Schwarzschild case. Considering the region inside the event horizon, where $\overline{r}<\overline{r}_H$, we have to invert the space-like and time-like characteristics, which implies $T^{0 }{}_{0}=p_{\|}$ and $T^{1}{}_{1}=-\rho$ \cite{Simpson:2018tsi}. Thus for this region we have

\begin{figure}[h!]
    \centering
    \includegraphics{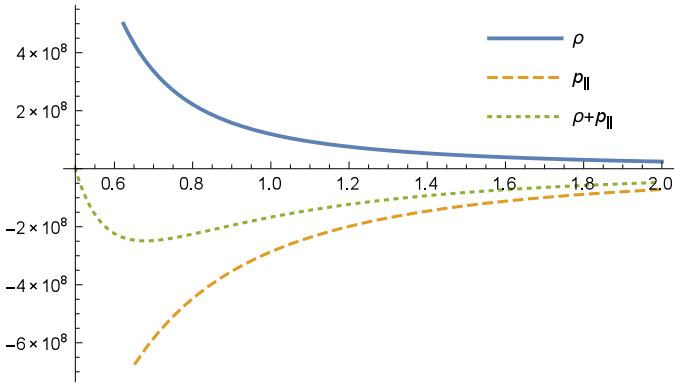}
    \caption{Energy density, radial pressure and $\rho + p_{\|}$ outside the event horizon}
    \label{fig1}
\end{figure}

\begin{equation}\label{36}
    \rho + p_{\|} = \frac{2a^2\alpha^2(\overline{r}^3-\overline{r}_H^3)}{8\pi G \overline{r}^5}, 
\end{equation}
which in this case is also always negative, which implies that the null energy condition violation occur in all cases. We can then generalize these results to both cases, inside and outside the event horizon as follows

\begin{equation}\label{37}
    \rho + p_{\|} = - \frac{2a^2\alpha^2|\overline{r}^3-\overline{r}_H^3|}{8\pi G \overline{r}^5}.
\end{equation}

\begin{figure}[h!]
    \centering
    \includegraphics{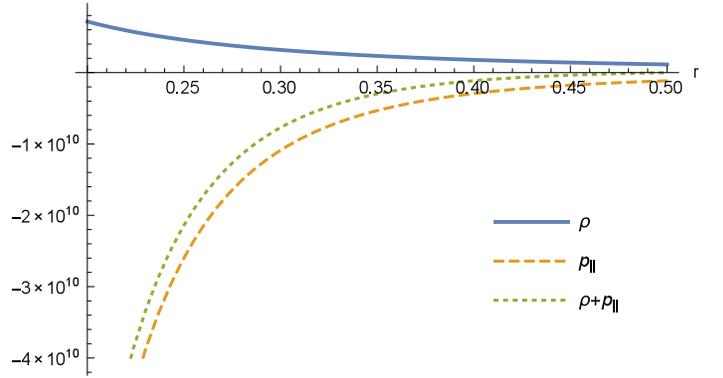}
    \caption{Energy density, radial pressure and $\rho + p_{\|}$ inside the event horizon}
    \label{fig2}
\end{figure}

The behaviour of the of the energy density, radial pressure and $\rho + p_{\|}$ outside and inside the event horizon is depicted in fig. (\ref{fig1}) and fig.(\ref{fig2}) respectively. The plots highlight that the null energy condition is always violated for both outside and inside event horizon regions. 

We conclude that the regularization of the black string metric follows the same pattern as the regularization of the Schwarzschild solution found by Simpson-Visser. The Simpson-Visser modified black string solution can generate a wormhole or black string depending on the parameter $a$ used in the metric, and its solutions are all regular if $a\neq 0$. However the energy conditions of the system are violated, that is, the type of regularization performed by Simpson-Visser results in the violation of the system conditions, analogous to what happens with the wormhole of Morris-Thorne \cite{Simpson:2018tsi, Morris:1988cz}.

\section{Applications of the regular solution of a black string}
Now, let us study the thermodynamic properties and the possible stable (or unstable) circular orbits for photons and massive particles. For such investigations we must consider the metric in the form (\ref{9}). It is important to note that, in this form, the position of the event horizon is different since $g_{00}$ has another form. Defining this position by $r_H$, we have

\begin{equation}\label{38}
    r_H = \sqrt{r_{HL}^2-a^2},
\end{equation}
where, as we have already seen, $r_{HL}=b^{1/3}/\alpha$.
\subsection{Regular black string thermodynamics}
The surface area bounded by the event horizon of a black hole always tends to increase or remain the same, which gives rise to the Second Law of Black Holes. This is analogous to the Second Law of Thermodynamics for entropy, as the entropy of a given system of particles cannot decrease either, ie $\delta S\geq 0$. We could imagine that this was just a coincidence or something superficial, since, in the case of the area of a black hole, this law has a mathematical rigor relative to General Relativity itself, while the second law of thermodynamics is not a law of nature, but rather a consequence of the fact that we are working statistically with a very large number of degrees of freedom in a physical system. However, it is possible to find a parallel between the laws of black holes and the other laws of thermodynamics, showing that this relationship is indeed something fundamental and not just a coincidence \cite{wald1984general}.



Given these thermodynamic relationships associated with the magnitudes of a black hole, we can then initially determine the temperature, known as the Hawking temperature, which can be given by the following expression:

\begin{equation}\label{39}
    T_{H}=\frac{\kappa}{2\pi}
\end{equation}
where $\kappa$ is surface gravity which is given by $-g_{00}'(r_H)/2$ \cite{Alencar:2018vvb}. So, we have

\begin{equation}\label{40}
    T_H=\frac{3\alpha^2}{4\pi}\sqrt{r_{HL}^2-a^2}=T_{HL}\sqrt{1-\frac{a^2}{r_{HL}^2}},
\end{equation}
where $T_{HL}=3\alpha^2r_{HL}/(4\pi)$ is the solution temperature of the usual black string \cite{Carvalho:2022eli}. Here we can see the same pattern observed in \cite{Simpson:2018tsi}, that is, we have a correction factor in relation to the non-regular solution of the type $\sqrt{1-a^2/k^2}$, where $ k=r_{HL}$. We see that, for $a=0$, we recover the usual black string solution, showing its consistency. When we have $a<r_{HL}$, the temperature tends to decrease with respect to $T_{HL}$, so that as $a\rightarrow r_{HL}$, $T_H$ tends to zero, such that this is achieved when we have the extreme throat wormhole case where $a=r_{HL}$. For $a>r_{HL}$, the result obtained in (\ref{40}) is no longer valid, which is reasonable, since in this interval we no longer have a black hole but a wormhole with no horizon of events.

\begin{figure}[h!]
    \centering
    \includegraphics[scale=0.9]{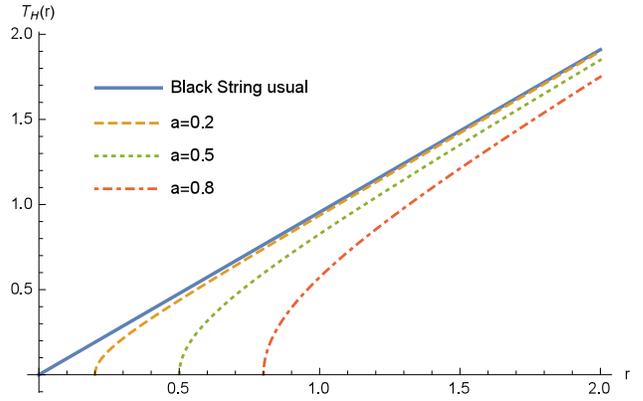}
    \caption{Hawking temperature}
    \label{fig3}
\end{figure}

In fig.(\ref{fig3}) we plot the behaviour of the Hawking temperature for three values of the bounce parameter $a$ and the usual black string case, for the sake of comparisson. As we can see, as we increase the value of the bounce parameter $a$ we also increase the value of $r_{HL}$ in which the Hawking temperature vanishes. Also, for large values of $r_{HL}$ we recover the usual linear behaviour of the black string Hawking temperature. 

Now one can also calculate the entropy and heat capacity, respectively, of the regular black string using the expressions $dS=dM/T$ and $C_V=dM/dT$ \cite{Alencar:2018vvb}. We can write the parameter $b$, defined in the black string metric, as $4M$ \cite{Rayimbaev:2021zxw}. So, as $b=\alpha^3(r_H^2+a^2)^{3/2}$, we have

\begin{equation}\label{41}
    dM=\frac{3}{4}\alpha^3r_H\sqrt{r_{H}^2+a^2}dr_H.
\end{equation}
Now, we can write $T_H$ as a function of $r_H$ and integrate the expression $dM/T$ to get the entropy as

\begin{equation}\label{42}
    S=S_L\left[\sqrt{1-\frac{a^2}{r_{HL}^2}}+\frac{a^2}{r_{HL}^2}\ln\left(\frac{\sqrt{r_{HL}^2-a^2}+r_{HL}}{a}\right)\right],
\end{equation}
where $S_L=(\alpha/2)\pi r_{HL}^2$ is the entropy of the usual black string \cite{Carvalho:2022eli}. Here we can see that the correction is not as trivial as in the case of Hawking temperature, because in addition to the factor $\sqrt{1-a^2/r_{HL}^2}$, we have a logarithmic function
 that depends on $r_{HL }$. First, we see that the solution is consistent with the non-regular solution, because when $a\rightarrow 0$, entropy tends to $S_L$. For $a<r_{HL}$, as $a$ approaches $r_{HL}$, the entropy value drops rapidly, such that at the limit $a\rightarrow r_{HL}$, entropy drops to zero. For $a>r_{HL}$, the entropy value is not valid.

\begin{figure}[h!]
    \centering
    \includegraphics[scale=0.9]{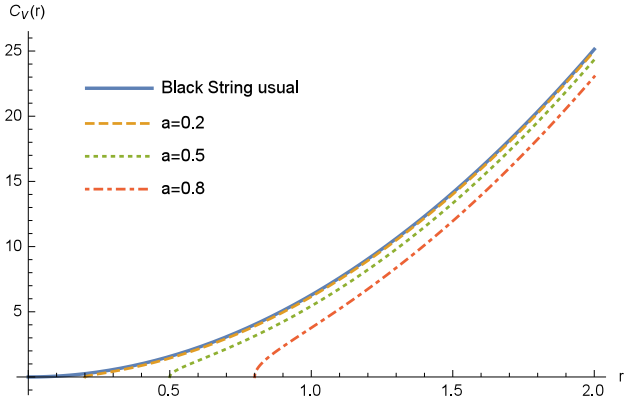}
    \caption{Heat capacity}
    \label{fig4}
\end{figure}

We can now calculate the heat capacity as:

\begin{equation}\label{43}
    C_V=\frac{dM}{dr_H}\frac{dr_H}{dT}=\alpha\pi r_H\sqrt{r_{H}^2+a^2} = C_{VL}\sqrt{1-\frac{a^2}{r_{HL}^2}},
\end{equation}
where $C_{VL}=\alpha\pi r_{HL}^2$ is the value of the heat capacity of the usual black string. We see that it follows exactly the same pattern as the temperature, that is, we have the factor $\sqrt{1-a^2/r_{HL}^2}$ correcting the usual value of the heat capacity, therefore, we can recover the usual result doing $a=0$, as expected. The sign of $C_V$ is important to evaluate the thermodynamic stability of the black string, because, if $C_V>0$, this implies that the black hole will be thermodynamically stable, and if $C_V<0$, we will have an unstable solution \cite{Carvalho:2022eli}. It can be seen in equation (\ref{43}) (and in fig.(\ref{fig4})) that $C_V$ will always be positive, as will $C_{VL}$, indicating that this solution will always be stable, with no possibility of evaporation, for example. 

Finally, let us evaluate the Helmholtz free energy, given by $F=M-T_HS$ \cite{Furtado:2022tnb}. Using the fact that $M=b/4$ and using the equations (\ref{40}) and (\ref{42}), we get:

\begin{eqnarray}\label{44}
    F=F_L\left(3\sqrt{1-\frac{a^2}{r_{HL}^2}}\frac{S}{S_L}-2\right),
\end{eqnarray}
where $F_L=-\alpha^3r_{HL}^3/8$ is the free energy of the usual solution. The result remains consistent, as when $a\rightarrow 0$, $S\rightarrow S_L$ and we retrieve the usual result $F_L$. Here, we also have a non-trivial dependence of the solution of $F$ on $a$ and $r_{HL}$, and when $a\rightarrow r_{HL}$, unlike the other quantities, $F$ does not tends to zero, but to $-2F_L$. Here we can see a pattern similar to what is found in BTZ-type black hole regularization \cite{Furtado:2022tnb}, that is, when $a$ is close to zero, the energy is negative, however, when $a$ approaches $r_{HL}$, the energy becomes positive, however, unlike \cite{Furtado:2022tnb}, free energy is not allowed for $a>r_{HL}$.

\subsection{Circular orbits in the regular black string}
An important result that we can find is the possible circular orbits for this solution. One way to assess this is by determining the effective potential energy ($V_{eff}$) of the system formed by the black string and a massive particle or a photon. A stable orbit is one in which, in the vicinity of the equilibrium point, the concavity of the function is positive, that is, $V_{eff}''>0$, and the unstable one, its concavity is negative, therefore, $V_{ eff}''<0$ \cite{Jefremov:2015gza}.

For the sake of symmetry, we will consider only the equatorial plane, that is, $z=0$. So that, we have

\begin{equation}\label{45}
        \left(\frac{dr}{d\tau}\right)^2=E^2+\left(\alpha^2(r^2+a^2)-\frac{b}{\alpha \sqrt{r^2+a^2}}\right)\left(\epsilon-\frac{L^2}{r^2+a^2}\right).
\end{equation}
We can establish that

\begin{equation}\label{46}
        \left(\frac{dr}{d\tau}\right)^2=E^2-V_{eff}(r),
\end{equation}
so, comparing the equation (\ref{45}) with (\ref{46}), we easily arrive at the expression for $V_{eff}$:

\begin{equation}\label{47}
        V_{eff}(r)=\alpha^2(r^2+a^2)\left(1-\frac{(r_H^2+a^2)^{3/2}}{ (r^2+a^2)^{3/2}}\right)\left(\frac{L^2}{r^2+a^2}-\epsilon \right).
\end{equation}

\begin{figure}[h!]
    \centering
    \includegraphics[scale=0.9]{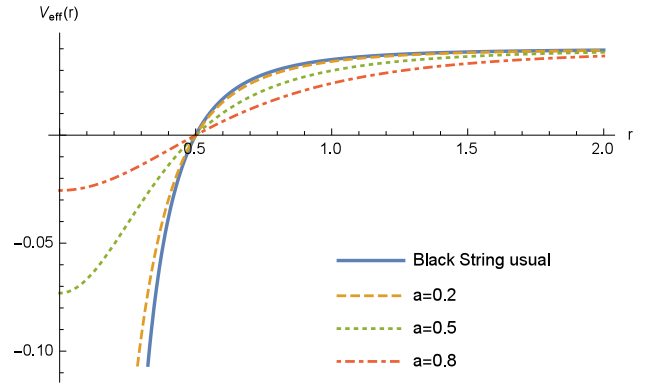}
    \caption{Effective potential for massless particles}
    \label{fig5}
\end{figure}

The first case that we will analyze is the circular path of light by imposing $\epsilon=0$ in (\ref{47}) and using the condition $V_{eff}'=0$. Therefore, we reach at the following results

\begin{equation}\label{48}
        V_{eff}(r)=\alpha^2L^2\left(1-\frac{(r_H^2+a^2)^{3/2}}{(r^2+a^2)^{3/2}}\right);
\end{equation}

\begin{equation}\label{49}
        V_{eff}'(r)=3\alpha^2L^2r\left(\frac{(r_H^2+a^2)^{3/2}}{ (r^2+a^2)^{5/2}}\right).
\end{equation}
Here we can see that the only possible solution for $V_{eff}'(r)=0$ is $r=0$. However, $r=0$ implies $\overline{r}=a$, that is, either the light would be in the throat of a wormhole, case $a\geq r_{HL}$, or it would be inside the horizon of events of a black hole, case $a<r_{HL}$. Therefore, it is clearly not a valid solution from a physical point of view \cite{Simpson:2018tsi}. Therefore, we can conclude that there are no circular orbits for a photon around a regular black string. The effective potential for massless particles is depicted in fig.(\ref{fig5}).

\begin{figure}[h!]
    \centering
    \includegraphics[scale=0.9]{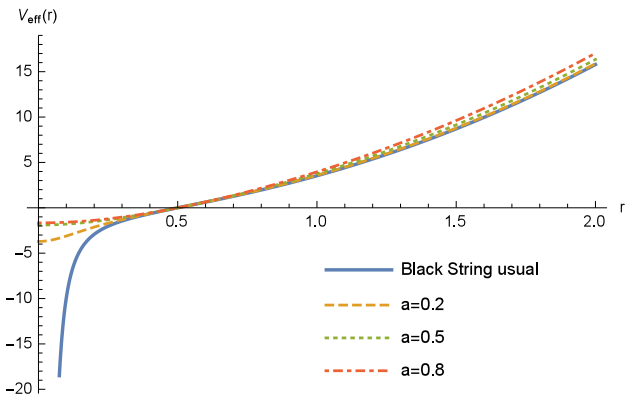}
    \caption{Effective potential for massive particles}
    \label{fig6}
\end{figure}

Let us now analyze the case of the massive particle. Using the same reasoning we used in the case of light, but using $\epsilon=-1$, we will get:

\begin{eqnarray}
        V_{eff}(r)&=&\alpha^2\left(1-\frac{(r_H^2+a^2)^{3/2}}{(r^2+a^2)^{3/2}}\right)(L^2+r^2+a^2); \label{50}\\
        \nonumber V_{eff}'(r)&=&\alpha^2r\frac{3(L^2+r^2+a^2)(r_H^2+a^2)^{3/2}}{(r^2+a^2)^{5/2}}+\\
        &&2\alpha^2r\left(1-\frac{(r_H^2+a^2)^{3/2}}{(r^2+a^2)^{3/2}}\right). \label{51} 
\end{eqnarray}
Imposing $V_{eff}'(r)=0$, assuming $r\neq 0$ and setting $L_c$ to be the angular momentum per unit mass of the circular orbit, and $r_c$ to be the radial coordinate value from the circular orbit, we can find $L_c$ as a function of $r_c$, $a$, and $r_H$:

\begin{equation}\label{52}
        L_c^2=-\left[\frac{(r_c^2+a^2)}{3}+\frac{2(r_c^2+a^2)^{5/2}}{3(r_H^2+a^2)^{3/2}}\right].
\end{equation}
Hence, it is seen that this expression cannot be physically valid, as the right side of the equation (\ref{52}) is always negative for any $r_c$, $a$, and $r_H$. As a consequence, we would have complex angular momentum, which is not physically allowed. Therefore, we cannot have circular orbits for massive particles around a regular black string either. These results are similar to the case of regularization of the BTZ black hole \cite{Furtado:2022tnb}, in which it is not possible to find circular orbits for either photons or massive particles, indicating a similarity between the two types of solutions, although well different from the case analyzed by Simpson-Visser, where orbits are found for the photon sphere and massive particles. The effective potential for massive particles is depicted in fig.(\ref{fig6}).

\section{Conclusion}

In this paper, we consider the solution already known in the literature of a black string, and we apply the same procedure employed by Simpson and Visser to regularize this solution. Thus, we obtained results similar to those found by Simpson-Visser, in which an interpolation between a regular black hole and a traversable wormhole is found, although this is not a Morris-Thorne wormhole type. The regularity of the solution could be verified by analyzing the tensors and curvature invariants, which are all finite at the origin since $a\neq 0$, where $a$ is the parameter used to regularize the solution.

After determining the solution and verifying its regularity, an analysis of the energy conditions related to the moment-energy tensor was also performed and once again we identified the same pattern of the Simpson-Visser regularization, that is, the null energy condition is violated inside and outside the event horizon for any kind of solution, whether regular black string or wormhole. 

After this analysis of curvature and energy conditions, we move on to basic applications of this solution. First, we analyzed the thermodynamic properties of the black string. We verified that all the calculated quantities, namely, the Hawking temperature, entropy, heat capacity, and Helmholtz free energy, undergo some change in relation to their respective values in the non-regular solution usual. In the case of the Hawking temperature and heat capacity, this change corresponds to the product of the usual value of the respective quantities with a correction factor of the type $\sqrt{1-a^2/k^2}$, where $k=r_{HL }=b^{1/3}/\alpha$ which is the position of the event horizon in the usual solution. In the case of entropy and Helmholtz energy, the correction is more complex, involving not only this factor, but also a logarithmic function. All quantities were consistent with the fact that when $a\rightarrow 0$, the usual solution is recovered. At the limit where $a\rightarrow r_{HL}$, with the exception of the Helmholtz free energy, all quantities tend to zero. In the case of free energy, it changes from a negative value in the usual solution to a positive value as it approaches $r_{HL}$. We must highlight also that the Simpson-Visser modification does not provide any change in the sign of the heat capacity, which is positive for any value of the bounce parameter $a$.

Finally, we analyze the possibilities of circular orbits for photons and massive particles. As a result, unlike the Simpson-Visser work, it was not possible to find circular orbits for the regular black string in any of the cases, since, for both photons and massive particles, the only solution is found in $r= 0$, which is not physically reasonable, as we would be inside the black hole or in the ``throat'' of a wormhole. Even if we try to find some value for the angular momentum that minimizes the circular orbit in the massive case, we find a negative value for $L_c^2$, that is, we would have a complex angular momentum as a solution, which is not physically allowed.

\section*{Acknowledgements}

G.A. would like to thank Conselho Nacional de Desenvolvimento Científico e Tecnológico (CNPq) and Fundação Cearense de Apoio ao Desenvolvimento Científico e Tecnológico
(FUNCAP) and A.L. would like to thank Coordenação de Aperfeiçoamento de Pessoal de Nível Superior - Brasil (CAPES) for finantial support

\end{document}